\documentclass[conference]{IEEEtran}
\usepackage{amssymb}
\usepackage{graphicx}
\usepackage{amsfonts}
\usepackage{mathrsfs}
\usepackage{cite,url}
\usepackage{upquote}
\usepackage{pifont}
\usepackage{amsmath}
\newcommand{\ra}{\rightarrow}
\usepackage{units}
\newcommand{\I}{\mathscr{I}}
\newcommand{\Ia}{\mathscr{I}_{\alpha}}
\newcommand{\X}{\mathbb{X}}

\newcommand{\R}{\mathbb{R}}

\newtheorem{thm}{Theorem}
\newtheorem{theorem}[thm]{Theorem}

\newtheorem{definition}[thm]{Definition}

\begin{document}
\title {Relative $\alpha$-Entropy Minimizers Subject to Linear Statistical Constraints}
\author{
\IEEEauthorblockN{M. Ashok Kumar}
\IEEEauthorblockA{Department of ECE \\
Indian Institute of Science \\
Bangalore, Karnataka 560012, India\\
Email: ashokm@ece.iisc.ernet.in}
\and
\IEEEauthorblockN{Rajesh Sundaresan}
\IEEEauthorblockA{Department of ECE\\
Indian Institute of Science\\
Bangalore, Karnataka 560012, India\\
Email: rajeshs@ece.iisc.ernet.in}
}

\maketitle

\begin{abstract}
We study minimization of a parametric family of relative entropies, termed relative $\alpha$-entropies (denoted $\mathscr{I}_{\alpha}(P,Q)$). These arise as redundancies under mismatched compression when cumulants of compressed lengths are considered instead of expected compressed lengths. These parametric relative entropies are a generalization of the usual relative entropy (Kullback-Leibler divergence). Just like relative entropy, these relative $\alpha$-entropies behave like squared Euclidean distance and satisfy the Pythagorean property. Minimization of $\mathscr{I}_{\alpha}(P,Q)$ over the first argument on a set of probability distributions that constitutes a linear family is studied. Such a minimization generalizes the maximum R\'{e}nyi or Tsallis entropy principle. The minimizing probability distribution (termed $\mathscr{I}_{\alpha}$-projection) for a linear family is shown to have a power-law.
\end{abstract}


\section{Introduction}
 \label{sec:introduction}

Maximum entropy principle is a well-known selection principle under uncertainty. This is an idea that dates back to L.~Boltzmann, was popularized by E.~T.~Jaynes \cite{1982xxPPSSP_Jay}, and has its foundation in the theory of large deviation. Suppose that an ensemble average measurement (say sample mean, sample second moment, or any other similar linear statistic) is made on the realization of a sequence of iid random variables. The realization must then have an empirical distribution that obeys the constraint placed by the measurement -- the empirical distribution must belong to an appropriate convex set, say $\mathbb{E}$. Large deviation theory tells us that a special member of $\mathbb{E}$, denoted $P^*$, is overwhelmingly more likely than the others. If the alphabet $\mathbb{X}$ is finite (with cardinality $|\mathbb{X}|$), and the prior probability distribution (before measurement) is the uniform distribution $U$ on $\mathbb{X}$, then $P^*$ is the one that minimizes the relative entropy\footnote{The relative entropy of $P$ with respect to $Q$ is defined as
\[
  \I(P \| Q) := \sum\limits_{x \in \X} P(x) \log \frac{P(x)}{Q(x)}
\]
and the Shannon entropy of $P$ is defined as
\[
 H(P) := - \sum\limits_{x \in \X} P(x) \log P(x).
\]
The usual convention is $p \log \frac{p}{q} = 0$ if $p = 0$ and $+\infty$ if $p > q = 0$.}
\[
  \mathscr{I}(P \| U) = \log |\mathbb{X}| - H(P),
\]
which is the same as the one that maximizes (Shannon) entropy\footnote{Hence the name maximum entropy principle.}, subject to $P \in \mathbb{E}$. In Jaynes' words, \emph{``... it is maximally noncommittal to the missing information''} \cite{1982xxPPSSP_Jay}.

As a physical example, let us tag a particular molecule in the atmosphere. Let $X$ denote the height of the molecule in the atmosphere. Then the potential energy of the molecule is $mgX$. Let us suppose that the average potential energy is held constant, that is, $E[mgX] = c$, a constant. Then the probability distribution of the height of the molecule is taken to be the exponential distribution $\lambda \exp{(-\lambda x)}$, where $\lambda = mg/c$. This is also the maximum entropy probability distribution subject to first moment constraint \cite{2006xxEIT_CovTho}. 

More generally, if the prior probability distribution (before measurement) is $Q$, then $P^*$ minimizes $\mathscr{I}(P \| Q)$ subject to $P \in \mathbb{E}$. Something more specific can be said: $P^*$ is the limiting conditional probability distribution of a ``tagged'' particle under the conditioning imposed by the measurement. This is called \emph{the conditional limit theorem} or the {\em Gibbs conditioning principle}; see for example Campenhout and Cover \cite{1981TIT_Cam_Cov} or Csisz\'ar \cite{1984xxAP_Csi} for a more general result.

It is well-known that $\mathscr{I}(P\|Q)$ behaves like ``squared Euclidean distance'' and has the ``Pythagorean property'' (Csisz\'ar \cite{1975xxAP_Csi}). In view of this and since $P^*$ minimizes $\mathscr{I}(P \| Q)$ subject to $P \in \mathbb{E}$, one says that $P^*$ is ``closest'' to $Q$ in the relative entropy sense amongst the probability distributions in $\mathbb{E}$, or in other words, ``$P^*$ is the $\mathscr{I}$-projection of $Q$ on $\mathbb{E}$''. Motivated by the above maximum entropy and Gibbs conditioning principles, $\mathscr{I}$-projection was extensively studied by Csisz\'{a}r \cite{1984xxAP_Csi}, \cite{1975xxAP_Csi}, and Csisz\'{a}r and Mat\'{u}\v{s} \cite{200306TIT_CsiMat}.

This paper is on the projection problem associated with a parametric generalization of relative entropy. To see how this parametric generalization arises, we return to our remark on how relative entropy arises in Shannon theory. For this, we must first recall how R\'enyi entropies are a parametric generalization of the Shannon entropy.

R\'{e}nyi entropies $H_{\alpha}(P)$ for $\alpha \in (0, 1)$ play the role of Shannon entropy when the \emph{normalized cumulant} of compression length,
\[
 \frac{1}{\rho}\log E[\exp\{\rho L(X)\}],
\]
is considered instead of expected compression length $E[L(X)]$, where $\rho >0$ is the cumulant parameter. Campbell \cite{1965xxIC_Cam} showed that the minimum normalized cumulant subject to all compression strategies that satisfy the Kraft inequality is 
\[
H_{\alpha}(P) = \frac{1}{1-\alpha}\log \sum_{x\in \X} P(x)^{\alpha},
\]
where $\alpha = 1/(1+\rho)$. We also have $\lim_{\alpha \to 1} H_{\alpha}(P) = H(P)$, so that R\'{e}nyi entropy may be viewed as a generalization of Shannon entropy.

If the compressor assumed that the true probability distribution is $Q$, instead of $P$, then the gap in the normalized cumulant's optimal value is an analogous parametric divergence quantity\footnote{Blumer and McEliece \cite{198809TIT_BluMcE}, in their attempt to find better upper and lower bounds on the redundancy of generalized Huffman coding, were indirectly bounding this parameterized divergence.}, which we shall denote $\mathscr{I}_{\alpha}(P,Q)$ \cite{200701TIT_Sun}. The same quantity also arises when we study the gap from optimality of mismatched guessing exponents. See Arikan \cite{199601TIT_Ari} and Hanawal and Sundaresan \cite{201101TIT_HanSun} for general results on guessing, and see Sundaresan \cite{200206ISIT_Sun} and \cite{200701TIT_Sun} on how $\mathscr{I}_{\alpha}(P,Q)$ arises in the context of mismatched guessing. Recently, Bunte and Lapidoth \cite{2014xxarx_BunLap} have shown that the $\mathscr{I}_{\alpha}(P,Q)$ also arises as redundancy in a mismatched version of the problem of coding for tasks.

$\mathscr{I}_{\alpha}$ may be expressed as
\begin{align}
\label{eqn:I-alpha}
 \lefteqn{\mathscr{I}_{\alpha}(P,Q)}\nonumber\\
& & = \frac{\alpha}{1-\alpha} \log \Big[ \sum_x P(x) Q(x)^{\alpha-1} \Big] - \frac{1}{1-\alpha}\log \sum_x P(x)^{\alpha}\nonumber\\
& & + \log \sum_x Q(x)^{\alpha}.
\end{align}
For each $\alpha >0, \alpha\neq 1$, $\Ia(P,Q)\ge 0$ with equality iff $P=Q$ \cite[Prop.~4]{200701TIT_Sun}. Also, $\mathscr{I}_{\alpha}(P,Q) = \infty$ only when either
\begin{itemize}
  \item $\alpha < 1$ and $\text{Supp}(P)\nsubseteq \text{Supp}(Q)$\footnote{$\text{Supp}(P)=\{x:P(x)>0\}$.} or
  \item $\alpha > 1$ and $\text{Supp}(P)\cap \text{Supp}(Q) = \emptyset$.
\end{itemize}
For $\alpha >1$, $\mathscr{I}_{\alpha}(P,Q)$ turns out to be relevant in a robust parameter estimation problem of statistics \cite{2008xxJma_FujEgu}.

As one might expect, it is known that (see for example, Sundaresan \cite[Sec.~V-5)]{200701TIT_Sun} or Johnson and Vignat \cite[A.1]{200705AIHP_JohVig}) $\lim_{\alpha \to 1} \mathscr{I}_{\alpha}(P,Q) = \mathscr{I}(P\|Q)$, so that we may think of relative entropy as $\mathscr{I}_1(P,Q)$. Thus $\mathscr{I}_{\alpha}$ is a generalization of relative entropy, i.e., a \emph{relative $\alpha$-entropy}\footnote{This terminology is from Lutwak, et al. \cite{200501TIT_LutYanZha}.}.

Not surprisingly, the maximum R\'{e}nyi entropy principle has been considered as a natural alternative to the maximum entropy principle of decision making under uncertainty. This principle is equivalent to another principle of maximizing the so-called Tsallis entropy which happens to be a monotone function of the R\'{e}nyi entropy. R\'{e}nyi entropy maximizers under moment constraints are distributions with a power-law decay (when $\alpha < 1$). See Costa et al. \cite{200307EMMCVPR_CosHerVig} or Johnson and Vignat \cite{200705AIHP_JohVig}. Many statistical physicists have studied this principle in the hope that it may ``explain'' the emergence of power-laws in many naturally occurring physical and socio-economic systems, beginning with Tsallis \cite{1988xxJSP_Tsa}. Based on our explorations of the vast literature on this topic, we feel that our understanding, particularly one that ought to involve a modeling of the \emph{dynamics} of such systems with the observed power-law profiles as equilibria in the asymptotics of large time, is not yet as mature as our understanding of the classical Boltzmann-Gibbs setting. But, by noting that $\mathscr{I}_{\alpha} (P,U) = \log |\mathbb{X}| - H_{\alpha}(P)$, we see that both the maximum R\'{e}nyi entropy principle and the maximum Tsallis entropy principle are particular instances of a ``minimum relative $\alpha$-entropy principle'': 
\[
 \text{ minimize } \mathscr{I}_{\alpha}(P,Q) \text{ over } P \in \mathbb{E}.
\]
We shall call the minimizing $P^*$ as the $\mathscr{I}_{\alpha}$-projection of $Q$ on $\mathbb{E}$.

The objective of this paper is to characterize the $\mathscr{I}_{\alpha}$-projection on a particular convex family called \emph{linear family} which we shall define now.

\vspace*{.1in}

\begin{definition}
\label{defn:linear-family}
A linear family characterized by $k$ functions $f_i: \X \ra \R$, $1 \leq i \leq k$, is the set of probability distributions given by
\begin{eqnarray}
\label{eqn:linear_family}
 \mathbb{L} = \Big\{P\in \mathcal{P}(\mathbb{X}): \sum\limits_x P(x)f_i(x) = 0, i = 1,\dots,k\Big\}.
\end{eqnarray}
\end{definition}

\vspace*{.1in}

We will show that the $\mathscr{I}_{\alpha}$-projection on a linear family comes from an $\alpha$-power law family analogous to the fact that the $\mathscr{I}_1$-projection on such a family comes from an exponential family \cite{2004xxITST_CsiShi}.

\section{Characterization of $\mathscr{I}_{\alpha}$-projection on a linear family}
\label{sec:projection}
In this section, we find the structure of the $\mathscr{I}_{\alpha}$-projection on a linear family $\mathbb{L}$ and prove a necessary and sufficient condition for a $P^*\in\mathbb{L}$ to be the $\mathscr{I}_{\alpha}$-projection on $\mathbb{L}$. We consider the cases $\alpha > 1$ and $\alpha < 1$ separately in the two subsections. The proof for the $\alpha <1$ case is similar to Csisz\'{a}r and Shields' proof for $\alpha =1$ case \cite{2004xxITST_CsiShi}. For the proof of $\alpha >1$ case, we will resort to the Lagrange multiplier technique.

\subsection{$\alpha <1$:}

The result for $\alpha <1$ is the following.

\vspace{0.1in}

\begin{figure}[tb]
\centering
 \includegraphics[height = 3cm, width = 4cm]{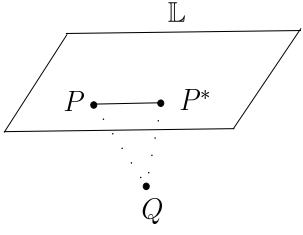}
\caption{\label{fig:pythagorean}Pythagorean property}
\end{figure}

\begin{theorem}
 \label{thm:projection_alpha<1}
Let $\alpha <1$. Let $\mathbb{L}$ be a linear family characterized by $f_i, i=1, \dots, k$. Let $Q$ be a probability distribution with $\text{Supp}(Q) = \mathbb{X}$. Then the following hold.
\begin{itemize}
 \item[(a)] $Q$ has an $\mathscr{I}_{\alpha}$-projection on $\mathbb{L}$. Call it $P^*$.

 \item[(b)] $\text{Supp}(P^*) = \text{Supp}(\mathbb{L})$\footnote{$\text{Supp}(\mathbb{L})$ is defined to be the union of supports of members of $\mathbb{L}$.} and the Pythagorean equality holds (see Figure \ref{fig:pythagorean}):
\begin{eqnarray}
 \label{eqn:pythagorean_equality}
  \mathscr{I}_{\alpha}(P, Q) = \mathscr{I}_{\alpha}(P, P^*) + \mathscr{I}_{\alpha}(P^* ,Q) \quad \forall P \in \mathbb{L}.
\end{eqnarray}

 \item[(c)] The $\mathscr{I}_{\alpha}$-projection $P^*$ satisfies 
\begin{multline}
 \label{eqn:power_law}
 P^*(x)  = Z^{-1} \cdot \Big[Q(x)^{\alpha -1} + (1-\alpha)\sum\limits_{i=1}^k \theta_i^* f_i(x)\Big]^{\frac{1}{\alpha -1}}\\
\forall x \in \mathbb{X},
\end{multline}
where $\theta_1^*, \dots, \theta_k^*$ are scalars and $Z$ is the normalization constant that makes $P^*$ a probability distribution.

 \item[(d)] The $\mathscr{I}_{\alpha}$-projection is unique. $\hfill \IEEEQEDopen$
\end{itemize}
\end{theorem}

\vspace{0.1in}

\begin{IEEEproof} (a) From (\ref{eqn:I-alpha}), it is clear that the mapping $P \mapsto \mathscr{I}_{\alpha}(P, Q)$ is continuous as each of the terms in (\ref{eqn:I-alpha}) is continuous in $P$.  Also $\mathbb{L}$ is compact. Hence the $\mathscr{I}_{\alpha}$-projection exists.

(b) Let $P\in \mathbb{L}$ and let $P_t = (1-t)P^*+tP$, $0 \le t \le 1$. Since $P_{t}\in \mathbb{L}$, the mean value theorem says that for each $t \in (0,1)$, there exists $\tilde{t}\in (0,t)$ such that
\begin{eqnarray}
\label{derivative1}
0\le \frac{1}{t}[\mathscr{I}_{\alpha}(P_{t},Q)-\mathscr{I}_{\alpha}(P^*,Q)] = \frac{d}{ds}\mathscr{I}_{\alpha}(P_{s},Q)|_{s=\tilde{t}}.
\end{eqnarray}
The first inequality follows from the fact that $P^*$ is the projection. Using (\ref{eqn:I-alpha}), we see that
\begin{eqnarray}
\label{derivative2}
\lefteqn{\frac{d}{ds}\mathscr{I}_{\alpha}(P_{s},Q) = \frac{\alpha}{1-\alpha} \left[ \frac{\sum\limits_x(P(x)-P^*(x))Q(x)^{\alpha-1}}{\sum\limits_x P_{s}(x)Q(x)^{\alpha-1}}\right.}\nonumber\\
& & \hspace{2cm} \left. - \frac{\sum\limits_x(P(x)-P^*(x))P_{s}(x)^{\alpha-1}}{\sum\limits_x P_{s}(x)^{\alpha}} \right].
\end{eqnarray}
As $t\downarrow 0$, from (\ref{derivative2}) and the inequality in (\ref{derivative1}), we get 
\begin{align}
\label{pythagorean_inequality_equivalent}
 \lefteqn{\frac{\sum\limits_x(P(x)-P^*(x))Q(x)^{\alpha-1}}{\sum\limits_x P^*(x)Q(x)^{\alpha-1}}}\nonumber\\
 & & \ge \frac{\sum\limits_x(P(x)-P^*(x))P^*(x)^{\alpha-1}}{\sum\limits_x P^*(x)^{\alpha}}.
\end{align}
That is,
\[
 \frac{\sum\limits_x P(x) Q(x)^{\alpha-1}}{\sum\limits_x P^*(x)Q(x)^{\alpha-1}} \ge \frac{\sum\limits_x P(x) P^*(x)^{\alpha-1}}{\sum\limits_x P^*(x)^{\alpha}},
\]
which, using (\ref{eqn:I-alpha}), can be seen to be equivalent to (\ref{eqn:pythagorean_equality}), but with inequality ``$\ge$'' in place of equality.

Suppose $P^*(x)=0$ for an $x \in \text{Supp}(P)$. Then $\alpha<1$ implies that right-hand side of (\ref{derivative2}) goes to $-\infty$ as $t\downarrow 0$, which contradicts the nonnegativity requirement in (\ref{derivative1}). Hence $\text{Supp}(P) \subseteq \text{Supp}(P^*)$. Since $P$ was arbitrary, we have $\text{Supp}(P^*) = \text{Supp}(\mathbb{L})$.

We will now establish equality in (\ref{eqn:pythagorean_equality}). Once again let $P\in \mathbb{L}$. Since $\text{Supp}(P^*) = \text{Supp}(\mathbb{L})$, we can find a new $\tilde t <0$ such that $P_t = (1-t)P^* + tP$ is a valid probability distribution for all $t\in (\tilde t, 0)$. Hence $P_t\in \mathbb{L}$ for all $t\in (\tilde t, 0)$. Since $t<0$, we have
\[
 0 \ge \frac{1}{t}[\mathscr{I}_{\alpha}(P_{t},Q)-\mathscr{I}_{\alpha}(P^*,Q)] \quad \forall t\in (\tilde t, 0).
\]
An argument similar to the one that led to (\ref{pythagorean_inequality_equivalent}) now proves that (\ref{pythagorean_inequality_equivalent}) holds with ``$\le$'' as well. This proves (\ref{eqn:pythagorean_equality}) and completes the proof of (b).

(c) This follows the proof of Csisz\'{a}r and Shields for case $\alpha =1$ \cite[Th.~3.4]{2004xxITST_CsiShi}.

From (\ref{eqn:linear_family}), it is clear that the probability distributions $P\in \mathbb{L}$, when considered as $|\text{Supp}(\mathbb{L})|$-dimensional vectors, belong to the orthogonal complement $\mathcal{F}^{\perp}$ of the subspace $\mathcal{F}$ of $\mathbb{R}^{|\text{Supp}(\mathbb{L})|}$ spanned by the vectors $f_i(\cdot), i = 1, \dots, k$, restricted to $\text{Supp}(\mathbb{L})$. These $P\in \mathbb{L}$ actually span $\mathcal{F}^{\perp}$. (This follows from the fact that if a subspace of $\mathbb{R}^{|\text{Supp}(\mathbb{L})|}$ contains a vector all of whose components are strictly positive, here $P^*$, then it is spanned by the probability vectors of that space.) Using (\ref{eqn:pythagorean_equality}), which is the same as (\ref{pythagorean_inequality_equivalent}) with equality, we have
\[
 \sum\limits_x P(x)\Bigg(\frac{Q(x)^{\alpha-1}}{\sum\limits_a P^*(a)Q(a)^{\alpha-1}} - \frac{P^*(x)^{\alpha-1}}{\sum\limits_a P^*(a)^{\alpha}}\Bigg) = 0 \quad \forall P\in \mathbb{L}.
\]
Consequently, the vector
\[
 \frac{Q(\cdot)^{\alpha-1}}{\sum\limits_a P^*(a)Q(a)^{\alpha-1}} - \frac{P^*(\cdot)^{\alpha-1}}{\sum\limits_a P^*(a)^{\alpha}}
\]
belongs to $(\mathcal{F}^{\perp})^{\perp} = \mathcal{F}$, that is,
\begin{align*}
 \frac{Q(x)^{\alpha-1}}{\sum\limits_a P^*(a)Q(a)^{\alpha-1}} - \frac{P^*(x)^{\alpha-1}}{\sum\limits_a P^*(a)^{\alpha}} = \sum_{i=1}^k \lambda_i f_i(x)\\ \forall x\in \text{Supp}(\mathbb{L}),
\end{align*}
for some scalars $\lambda_i, i=1, \dots, k$ . This verifies (\ref{eqn:power_law}) for obvious choices of $Z$ and $\theta_i^*, i=1, \dots, k$.

(d) Let $P^*_1$ and $P^*_2$ be two projections of $Q$ on $\mathbb{L}$. Then $\mathscr{I}_{\alpha}(P^*_1,Q)=\mathscr{I}_{\alpha}(P^*_2,Q)$. By (c), we have
\[
  \mathscr{I}_{\alpha}(P^*_2,Q) = \mathscr{I}_{\alpha}(P^*_2,P^*_1) + \mathscr{I}_{\alpha}(P^*_1,Q).
\]
Canceling $\mathscr{I}_{\alpha}(P^*_2,Q)$ and $\mathscr{I}_{\alpha}(P^*_1,Q)$, we get $\mathscr{I}_{\alpha}(P^*_2,P^*_1) = 0$ which implies $P^*_1 = P^*_2$.

\end{IEEEproof}

\vspace{0.05in}

One can also have a converse.

\vspace{0.1in}

\begin{theorem}
\label{thm:projection_alpha<1_converse}
 Let $\alpha <1$. Let $P^* \in \mathbb{L}$ be a probability distribution of the form (\ref{eqn:power_law}). Then 
$P^*$ satisfies (\ref{eqn:pythagorean_equality}) and $P^*$ is the $\mathscr{I}_{\alpha}$-projection $Q$ on $\mathbb{L}$. $\hfill \IEEEQEDopen$
\end{theorem}

\vspace{0.1in}

\begin{IEEEproof}
 If $P^*\in \mathbb{L}$ is of the stated form, then since every $P\in \mathbb{L}$ satisfies
\[
\sum\limits_{i=1}^k P(x) f_i(x) = \sum\limits_{i=1}^k P^*(x)f_i(x) = 0, \quad i=1, \dots, k,
\]
we have from (\ref{eqn:power_law}) that
\[
 Z^{\alpha-1} \sum\limits_x P(x) P^*(x)^{\alpha-1} =  \sum\limits_{x} P(x) Q(x)^{\alpha-1},
\]
and
\[
 Z^{\alpha-1} \sum\limits_x P^*(x)^{\alpha} = \sum\limits_{x} P^*(x) Q(x)^{\alpha-1}.
\]
Combining the above two equations to eliminate $Z^{\alpha-1}$, we get
\[
 \sum\limits_x P(x) Q(x)^{\alpha-1} = \frac{\sum\limits_x P^*(x)Q(x)^{\alpha-1}}{\sum\limits_x P^*(x)^{\alpha}}\sum\limits_x P(x) P^*(x)^{\alpha-1},
\]
which, using (\ref{eqn:I-alpha}), can be seen to be equivalent to (\ref{eqn:pythagorean_equality}). Thus, for any $P\in \mathbb{L}$, we have
\begin{eqnarray*}
  \mathscr{I}_{\alpha}(P,Q) & = & \mathscr{I}_{\alpha}(P,P^*) + \mathscr{I}_{\alpha}(P^*,Q)\\
& \ge & \mathscr{I}_{\alpha}(P^*,Q),
\end{eqnarray*}
which implies that $P^*$ is the $\mathscr{I}_{\alpha}$-projection of $Q$ on $\mathbb{L}$.
\end{IEEEproof}

\vspace{0.1in}

When $\alpha >1$, in general, $\text{Supp}(P^*) \neq \text{Supp}(\mathbb{L})$ as shown by the following counterexample.

\vspace{0.1in}

 Take $\alpha = 2$ and $\mathbb{X} = \{ 1,2,3,4 \}$. Take $Q=(\nicefrac{1}{4},\nicefrac{1}{4},\nicefrac{1}{4},\nicefrac{1}{4})$. Consider the linear family,
\begin{eqnarray*}
\mathbb{L} & = & \{P \in \mathcal{P}(\mathbb{X}): 8p_1 + 4p_2 + 2p_3 + p_4 = 7\}.
\end{eqnarray*}
Thus
\[
 \mathbb{L} = \Big\{P \in \mathcal{P}(\mathbb{X}): \sum\limits_x P(x)f_1(x) = 0 \Big\},
\]
where $f_1(\cdot) = (1,-3,-5,-6)$.

We claim that the $\mathscr{I}_{\alpha}$-projection of $Q$ on $\mathbb{L}$ is $P^* = (\nicefrac{3}{4}, \nicefrac{1}{4}, 0, 0)$. To check this claim, first note that $P^*\in \mathbb{L}$. Also, with $Z=\nicefrac{2}{5}$ and $\theta_1^* = -\nicefrac{1}{20}$, we can check that
\begin{eqnarray*}
  0 & < & Z P^*(x) = Q(x) + \theta_1^* f_1(x),\, x=1,2,\\
  0 & = & Z P^*(3) = Q(3) + \theta_1^* f_1(3),\\
  0 & = & Z P^*(4) > Q(4) + \theta_1^* f_1(4).  
\end{eqnarray*}
Hence, for any $P\in \mathbb{L}$,
\begin{eqnarray}
\label{eqn:inequality}
 Z^{\alpha-1} \sum\limits_{x=1}^4 P(x) P^*(x)^{\alpha-1}
 & = & Z \sum\limits_{x=1}^4 P(x) P^*(x)\nonumber\\
 & \ge & \sum\limits_{x=1}^4 P(x) [Q(x) + \theta_1^* f_1(x)]\nonumber\\
 & = & \sum\limits_{x=1}^4 P(x) Q(x)\nonumber\\
 & = & \sum\limits_{x=1}^4 P(x) Q(x)^{\alpha-1},
\end{eqnarray}
where the penultimate equality follows because $P\in \mathbb{L}$. Similarly, one can show that
\begin{equation*}
 Z^{\alpha -1} \sum\limits_x P^*(x)^{\alpha} = \sum\limits_x P^*(x)Q(x)^{\alpha -1}.
\end{equation*}
Combining this with (\ref{eqn:inequality}) to eliminate $Z^{\alpha -1}$, we get
\[
\sum\limits_x P(x)Q(x)^{\alpha -1} \le \frac{\sum\limits_x P^*(x)Q(x)^{{\alpha-1}}}{\sum\limits_x P^*(x)^{\alpha}} \sum\limits_x P(x)P^*(x)^{{\alpha-1}},
\]
which, using (\ref{eqn:I-alpha}), can be seen to be equivalent to
\begin{eqnarray}
\label{eqn:pythagorean_inequality1}
 \mathscr{I}_{\alpha}(P,Q)
 & \ge & \mathscr{I}_{\alpha}(P,P^*) + \mathscr{I}_{\alpha}(P^*,Q)\\
 & \ge & \mathscr{I}_{\alpha}(P^*,Q)\nonumber.
\end{eqnarray}
Thus, $P^*$ is the $\mathscr{I}_{\alpha}$-projection of $Q$ on $\mathbb{L}$.

Clearly $\text{Supp}(P^*) \subsetneqq \text{Supp}(\mathbb{L})$. Also for $P = (0.8227,0.0625,0.0536,0.0612)\in\mathbb{L}$, numerical calculations yield a strict inequality in (\ref{eqn:pythagorean_inequality1}) since the left-hand side and the right-hand side of (\ref{eqn:pythagorean_inequality1}) evaluate to $1.0114$ and $0.9871$, respectively.

As a consequence of this, the proof of Theorem \ref{thm:projection_alpha<1} for the case $\alpha <1$ cannot be carried forward to establish the structure of $\Ia$-projection for the case $\alpha >1$. We will resort to the usual Lagrange multiplier technique for this case.

\vspace{0.1in}

 \subsection{$\alpha >1$:}
   \label{sec:projection_alpha>1}

   We now establish the form of the $\mathscr{I}_{\alpha}$-projection on a linear family when $\alpha >1$.

\vspace{0.1in}

\begin{theorem}
 \label{thm:projection_alpha>1}
Let $\alpha >1$. Let $\mathbb{L}$ be a linear family characterized by $f_i,i=1, \dots, k$. Let $Q$ be a probability distribution with $\text{Supp}(Q) = \mathbb{X}$. Then the following hold.
\begin{itemize}
 \item[(a)] $Q$ has an $\mathscr{I}_{\alpha}$-projection on $\mathbb{L}$. Call it $P^*$.

 \item[(b)] The $\mathscr{I}_{\alpha}$-projection $P^*$ satisfies 
\begin{multline}
 \label{eqn:power_law_with+}
P^*(x) =  Z^{-1} \cdot \Big[Q(x)^{\alpha -1} + (1-\alpha)\sum\limits_{i=1}^k \theta_i^* f_i(x)\Big]_{+}^{\frac{1}{\alpha - 1}}\\
\forall x \in \mathbb{X},
\end{multline}
where $\theta_1^*, \dots, \theta_k^*$ are scalars and $Z$ is the normalization constant that makes $P^*$ a probability distribution and $[u]_{+} = \max\{u,0\}$.

\item[(c)] The Pythagorean inequality holds:
\begin{eqnarray}
 \label{eqn:pythagorean_inequality2}
  \mathscr{I}_{\alpha}(P,Q) \ge \mathscr{I}_{\alpha}(P,P^*) + \mathscr{I}_{\alpha}(P^*,Q) \quad \forall P \in \mathbb{L}.
\end{eqnarray}

\item[(d)] The $\mathscr{I}_{\alpha}$-projection is unique.

\item[(e)] If $\text{Supp}(P^*) = \text{Supp}(\mathbb{L})$, then (\ref{eqn:pythagorean_inequality2}) holds with equality.
\end{itemize}
$\hfill \IEEEQEDopen$
\end{theorem}

\vspace{0.1in}

\begin{IEEEproof}
(a) Same as proof of Theorem \ref{thm:projection_alpha<1}-(a).

(b) The optimization problem for the $\mathscr{I}_{\alpha}$-projection is
\begin{align}
\min_P \, & \mathscr{I}_{\alpha}(P,Q)\label{min}\\
\mbox{subject to } &\sum\limits_x P(x)f_i(x) = 0, \quad i=1, \dots, k \label{eqn:linear_constraints}\\
                   &\sum\limits_x P(x)       = 1 \label{eqn:probability_constraint}\\
                   &P(x)                    \ge 0 \quad \forall x \in \mathbb{X}. \label{eqn:positivity_constraints}
\end{align}
We will proceed in a sequence of steps.
\begin{itemize}
 \item[(i)] Observe that $\mathscr{I}_{\alpha}(\cdot,Q)$, in addition to being continuous, is also continuously differentiable. Indeed, we have
\begin{align}
\label{eqn:partial_derivative}
 \lefteqn{\frac{\partial}{\partial P(x)}\mathscr{I}_{\alpha}(P,Q)}\nonumber\\
 & & = \frac{\alpha}{1-\alpha}\Bigg[\frac{Q(x)^{\alpha - 1}}{\sum\limits_a P(a)Q(a)^{{\alpha-1}}} - \frac{P(x)^{\alpha - 1}}{\sum\limits_a P(a)^{\alpha}}\Bigg].
\end{align}
Both denominators are bounded away from zero because for any $P \in \mathbb{L}$, we have $\max_x P(x) \ge 1/|\mathbb{X}|$, and therefore 
\[
 \sum\limits_a P(a)Q(a)^{\alpha-1} \ge \frac{1}{|\mathbb{X}|}\cdot \min_a Q(a)^{\alpha-1} >0,
\]
and
\[
 \sum\limits_a P(a)^{\alpha} \ge \frac{1}{|\mathbb{X}|^{\alpha}} >0.
\]
Consequently, the partial derivative (\ref{eqn:partial_derivative}) exists everywhere on $\mathbb{R}_{+}^{|\mathbb{X}|}$, and is continuous because the terms involved are continuous. (The numerator of the second term in (\ref{eqn:partial_derivative}) is continuous because $\alpha >1$).
 \item[(ii)] Since the equality constraints in (\ref{eqn:linear_constraints}) and (\ref{eqn:probability_constraint}) arise from affine functions, and the inequality constraints in (\ref{eqn:positivity_constraints}) arise from linear functions, we may apply \cite[Prop.~3.3.7]{2003xxNLP_Ber} to conclude that there exist Lagrange multipliers ($\lambda_i, i=1, \dots, k$), $\nu$, and $\left(\mu(x),x\in \mathbb{X}\right)$ associated with the constraints (\ref{eqn:linear_constraints}), (\ref{eqn:probability_constraint}), and (\ref{eqn:positivity_constraints}), respectively, that satisfy:
\begin{eqnarray}
\label{eqn:lagrange1}
 \lefteqn{\frac{\alpha}{1-\alpha} \Bigg[\frac{P^*(x)^{\alpha - 1}}{\sum\limits_a P^*(a)^{\alpha}} - \frac{Q(x)^{\alpha - 1}}{\sum\limits_a P^*(a)Q(a)^{\alpha-1}}\Bigg]}\nonumber\\
& &  \hspace{1.5cm} = \sum\limits_{i=1}^k \lambda_i f_i(x) - \mu(x) + \nu \quad \forall x\\
 \label{eqn:feasibility_condition}
& & \hspace{0.7cm} \mu(x) \ge 0 \quad \forall x\\
\label{eqn:slackness_condition}
& & \mu(x)P^*(x) = 0 \quad \forall x.
\end{eqnarray}
In writing (\ref{eqn:lagrange1}), we have substituted (\ref{eqn:partial_derivative}) for $\frac{\partial}{\partial P(x)}\mathscr{I}_{\alpha}(P,Q)$.

\item[(iii)] Multiplying (\ref{eqn:lagrange1}) by $P^*(x)$, summing over all $x\in \mathbb{X}$, using $P^*\in \mathbb{L}$, and (\ref{eqn:slackness_condition}), we see that $\nu = 0$.

\item[(iv)] If $P^*(x) >0$, we must have $\mu(x) =0$ from (\ref{eqn:slackness_condition}), and its substitution in (\ref{eqn:lagrange1}) yields, for all such $x$,
\begin{align}
\label{eqn:lagrange2}
 \lefteqn{\frac{P^*(x)^{\alpha - 1}}{\sum\limits_a P^*(a)^{\alpha}}}\nonumber\\
& & = \frac{Q(x)^{\alpha - 1}}{\sum\limits_a P^*(a)Q(a)^{\alpha-1}} + \frac{1-\alpha}{\alpha}\sum\limits_{i=1}^k \lambda_i f_i(x).
\end{align}
If $P^*(x) =0$, (\ref{eqn:lagrange1}) implies that
\begin{eqnarray}
\label{eqn:lagrange3}
 \lefteqn{\frac{Q(x)^{\alpha - 1}}{\sum\limits_a P^*(a)Q(a)^{\alpha-1}} + \frac{1-\alpha}{\alpha}\sum\limits_{i=1}^k \lambda_i f_i(x)}\nonumber\\
 & & \hspace{3cm} = \frac{(1-\alpha)}{\alpha} \mu(x)\nonumber\\
 & & \hspace{3cm} \le 0,
\end{eqnarray}
where the last inequality holds because of (\ref{eqn:feasibility_condition}) and $\alpha >1$.
Therefore, (\ref{eqn:lagrange2}) and (\ref{eqn:lagrange3}) may be combined as 
\begin{multline*}
Z^{\alpha -1} P^*(x)^{\alpha - 1}  =  \Big[Q(x)^{\alpha -1} + (1-\alpha)\sum\limits_{i=1}^k \theta_i^* f_i(x)\Big]_{+}\\
\forall x \in \mathbb{X},
\end{multline*}
for obvious choices of $Z$ and $\theta_i^*$. This verifies (\ref{eqn:power_law_with+}) and completes the proof of (b).
\end{itemize}

(c) Using (\ref{eqn:power_law_with+}), for any $P \in \mathbb{L}$, we have
\begin{eqnarray}
\label{eqn:expected_value1}
 \lefteqn{\sum\limits_x P(x)P^*(x)^{{\alpha-1}}}\nonumber\\
 & & \ge Z^{1-\alpha} \sum\limits_x P(x)\Big[Q(x)^{\alpha -1} + (1-\alpha)\sum\limits_{i=1}^k \theta_i^* f_i(x)\Big]\nonumber\\
 & & = Z^{1-\alpha} \sum\limits_x P(x)Q(x)^{\alpha -1},
\end{eqnarray}
where the first inequality follows because the terms on the right-hand side corresponding to $x$ with $P^*(x)=0$ are nonpositive, and the last equality follows because $P\in \mathbb{L}$. Similarly,
\[
 \sum\limits_x P^*(x)^{\alpha} = Z^{1-\alpha} \sum\limits_x P^*(x)Q(x)^{\alpha -1}.
\]
Combining the above and (\ref{eqn:expected_value1}) we get
\begin{equation}
 \label{eqn:pyth_ineq_simplified}
\sum\limits_x P(x)Q(x)^{\alpha -1} \le \frac{\sum\limits_x P^*(x)Q(x)^{{\alpha-1}}}{\sum\limits_x P^*(x)^{\alpha}} \sum\limits_x P(x)P^*(x)^{{\alpha-1}},
\end{equation}
which, using (\ref{eqn:I-alpha}), is equivalent to (\ref{eqn:pythagorean_inequality2}). This completes the proof of (c).

(d) Same as proof of Theorem \ref{thm:projection_alpha<1}-(d).

(e) If $\text{Supp}(P^*) = \text{Supp}(\mathbb{L})$, then we have equality in (\ref{eqn:expected_value1}), hence equality in (\ref{eqn:pyth_ineq_simplified}), and hence equality in (\ref{eqn:pythagorean_inequality2}). This concludes the proof of (e) and the theorem.
\end{IEEEproof}

\vspace{0.1in}

As in the $\alpha <1$ case, one has a converse.

\vspace{0.1in}

\begin{theorem}
\label{thm:projection_alpha>1_converse}
 Let $\alpha >1$. Let $P^* \in \mathbb{L}$ be a probability distribution of the form (\ref{eqn:power_law_with+}). Then 
$P^*$ satisfies (\ref{eqn:pythagorean_inequality2}) for every $P\in \mathbb{L}$, and $P^*$ is the $\mathscr{I}_{\alpha}$-projection $Q$ on $\mathbb{L}$. $\hfill \IEEEQEDopen$
\end{theorem}

\vspace{0.1in}

\begin{IEEEproof}
The proof of Theorem \ref{thm:projection_alpha>1}-(c) shows that (\ref{eqn:pythagorean_inequality2}) holds. Now, for any $P\in \mathbb{L}$, we have
\begin{eqnarray*}
  \mathscr{I}_{\alpha}(P,Q) & \ge & \mathscr{I}_{\alpha}(P,P^*) + \mathscr{I}_{\alpha}(P^*,Q)\\
& \ge & \mathscr{I}_{\alpha}(P^*,Q),
\end{eqnarray*}
which implies that $P^*$ is the $\mathscr{I}_{\alpha}$-projection $Q$ on $\mathbb{L}$.
\end{IEEEproof}

\section{Concluding remarks}
Motivated by the maximum entropy principle, we studied minimization of a parametric extension of relative entropy, where the minimization was with respect to the first argument. We recognize that the minimizer on a set determined by linear statistical constraints belongs to the $\alpha$-power-law family. This is analogous to the fact that the minimizer of relative entropy ($\alpha =1$) belongs to the exponential family. A complementary minimization problem, where the minimization is with respect to the second argument of $\mathscr{I}_{\alpha}$, is relevant in robust statistics ($\alpha >1$) and in constrained compression settings ($\alpha <1$). This problem and the connection between the two minimization problems will be the subject matter of our forthcoming work.

\section*{Acknowledgments}
M. Ashok~Kumar was supported by a Council for Scientific and Industrial Research (CSIR) fellowship and by the Department of Science and Technology. R. Sundaresan was supported in part by the University Grants Commission by Grant Part (2B) UGC-CAS-(Ph.IV) and in part by the Department of Science and Technology.

\bibliographystyle{IEEEtran}
{
\bibliography{IEEEabrv,NCCFEB2015}
}

\end{document}